\documentclass[reprint,aps,prper,twocolumn]{revtex4-2} 
\usepackage{graphicx}
\usepackage{hyperref}
\usepackage{footnote}
\usepackage{amsmath,amssymb}
\usepackage{multirow}
\usepackage{tabularx}
\usepackage[table]{xcolor}
\usepackage{booktabs} 
\usepackage{longtable}
\usepackage{cleveref}
\usepackage{makecell}
\usepackage{array}
\usepackage{supertabular}
\usepackage{comment}
\usepackage{enumitem} 
\newcommand{\rowgroup}[1]{\hspace{-1em}#1}

\usepackage{enumitem}

\begin{document}

\title{Implementation and goals of quantum optics experiments in undergraduate instructional labs}
\date{\today}

\author{Victoria Borish}
\email[]{victoria.borish@colorado.edu}
\author{H. J. Lewandowski}

\affiliation{Department of Physics, University of Colorado, Boulder, Colorado 80309, USA}
\affiliation{JILA, National Institute of Standards and Technology and University of Colorado, Boulder, Colorado 80309, USA}

\begin{abstract}

As quantum information science and technology (QIST) is becoming more prevalent and occurring not only in research labs but also in industry, many educators are considering how best to incorporate learning about quantum mechanics into various levels of education. Although much of the focus has been on quantum concepts in non-lab courses, current work in QIST has a substantial experimental component. Many instructors of undergraduate lab courses want to provide their students the opportunity to work with quantum experiments. One common way this is done is through a sequence of quantum optics experiments often referred to as the ``single-photon experiments.'' These experiments demonstrate fundamental quantum phenomena with equipment common to research labs; however, they are resource intensive and cannot be afforded by all institutions. It is therefore imperative to know what unique affordances these experiments provide to students. As a starting point, we surveyed and interviewed instructors who use the single-photon experiments in undergraduate courses, asking how and why they use the experiments. We describe the most commonly used experiments in both quantum and beyond-first-year lab courses, the prevalence of actions the students perform, and the learning goals, ranging from conceptual knowledge to lab skills to student affect. Finally, we present
some strategies from these data demonstrating how instructors have addressed the common challenges of preparing students to work with conceptually and technically complex experiments and balancing the practice of technical skills with the completion of the experiments.

\end{abstract}

\maketitle

\section{Introduction}\label{sec:intro}

Quantum information science and technology (QIST) is a burgeoning field, in which research is now being conducted not only in academic and national laboratories, but also in industry. In order to help prepare students for the myriad of opportunities offered by this growing field, educators are beginning to create many new quantum education programs \cite{asfaw2022building, perron2021quantum, meyer2022todays}. To understand the new educational needs, education researchers are taking various approaches, including asking quantum industry what skills they value in potential new hires \cite{fox2020preparing, hughes2022assessing, hasanovic2022quantum, greinert2022future} and cataloguing the current state of QIST courses \cite{plunkett2020survey,cervantes2021overview, perron2021quantum, aiello2021achieving, kaur2022defining,meyer2022todays}. 
Much of the work so far has focused on improving conceptual learning in typical lecture-based courses \cite{singh2008interactive, mckagan2008developing,singh2015review} where the focus has been shifting from descriptions of possible ways to restructure course content to empirical studies of the teaching and learning of quantum physics \cite{bitzenbauer2021quantum}. Providing students opportunities to perform quantum experiments may have additional benefits \cite{borish2022seeing}, but there have been very few studies to-date investigating best practices for teaching quantum lab experiments.

One set of quantum optics experiments, in particular, has been gaining popularity in the advanced labs community in the United States over the past two decades. This sequence of experiments uses heralded and entangled photons to demonstrate various quantum phenomena such as single-photon interference and a violation of local realism. These experiments are commonly referred to as the ``single-photon experiments,'' and we will be referring to them as such throughout this work. The prevalence of these experiments is due, in part, to the efforts of the Advanced Laboratory Physics Association (ALPhA), a 501(c)(3) organization focused on fostering communication between instructors of intermediate and advanced lab courses at colleges and universities \cite{alphaWebsite}.  Between 2010 and 2021, ALPhA sponsored workshops (Immersions) attended by over 100 instructors in total, teaching them how to implement the experiments. A similar number of instructors have purchased some of the required equipment through ALPhA during that time. There is even a waitlist of instructors waiting to purchase discounted equipment in order to set up these experiments.

The single-photon experiments are very resource expensive, and it has not yet been clearly demonstrated whether the cost is worth the benefits to the students. The experimental setup requires a substantial amount of money for the equipment, instructor time to learn and set up the experiments, and instructor expertise to work with the apparatus. One of these setups costs between \$20,000 -- \$30,000 to purchase individual components that must be assembled \cite{markWebsite} and tens of thousands of dollars more for pre-built setups \cite{Qubitekk,qutools}. The cost is an insurmountable obstacle for many institutions. Nonetheless, many instructors are very excited about these experiments, and there are numerous articles describing the experiments and theory behind them with a pedagogical lens (see, for example, Refs.~\cite{galvez2005interference, pearson2010hands, beck2012quantum}). However, student experiences working with these experiments have so far only been briefly studied alongside reports of how these experiments are used in specific programs \cite{lukishova2022fifteen,galvez2010qubit,pearson2010hands}. It is important for education researchers to study these experiments in a wider variety of contexts to understand which parts of the experiments are necessary for certain learning goals, and which learning goals may be achieved with less expensive equipment. This knowledge could lead to the development of lower cost options to enable all institutions to provide their students the ability to attain certain learning goals, which is particularly relevant given the current pressure to include quantum in the curriculum \cite{nationalQuantumInitiativeAct}. 

In order to study the efficacy of the single-photon experiments, it is necessary to first understand what is the intended purpose of incorporating the experiments in undergraduate courses and how students are using them. We therefore aim to answer the following research questions:
\begin{enumerate} [label=RQ\arabic*:]
    \item How are the single-photon experiments incorporated into the undergraduate curriculum?
    \item What learning goals do instructors have for these experiments?
\end{enumerate}
To answer these questions, we conducted a survey and follow-up interviews of instructors who use the single-photon experiments in their courses. This study is a necessary first step for researchers wanting to study these experiments in the future, as it illustrates the landscape of the use and goals of the experiments, which is needed to formulate research questions and methodologies to study student learning with the experiments. Additionally, the results presented here are useful for instructors because they provide examples of a variety of strategies for using these experiments, something requested by several of the instructors we interviewed.

\section{Background}

The single-photon experiments are complicated to study because they bring together many different facets, which we briefly describe in this section. We begin by detailing the single-photon experiments themselves, including a description of the experimental setup, the variety of different possible experiments that can be done with similar setups, and previous work published about educational uses of this sequence of experiments. We then discuss prior work related to some of the possible learning goals of the single-photon experiments. This includes various ways instructors teach about the concepts covered in these experiments, such as by utilizing simulations or videos of real experiments. We additionally present a variety of non-conceptual learning goals of many beyond-first-year (BFY) lab courses, since this is the kind of course in which the single-photon experiments are most commonly used (see Sec.~\ref{sec:courses}). This will provide context when interpreting the results and determining what is unique about the single-photon experiments.

\subsection{Single-photon experiments}

\begin{figure}[htbp]
  \centering
  \includegraphics[width=\columnwidth]{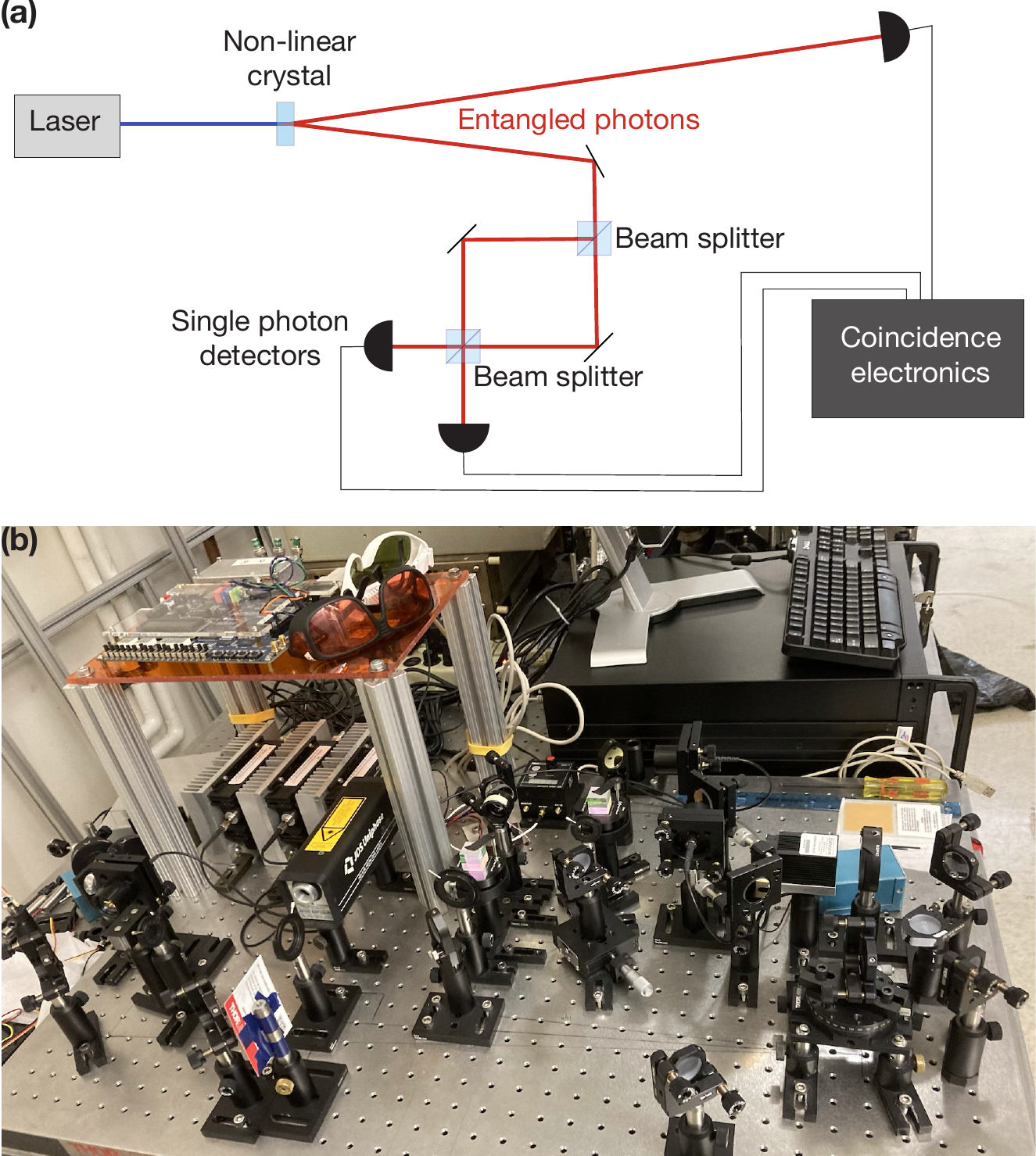}
\caption{The (a) schematic of the experiment and (b) photo of an actual experiment for the single-photon interferometer.}
\label{fig:experimentalSetup}
\end{figure}

Although there are slight variations in the experimental setups for each specific experiment, all of the single-photon experiments begin with a typically 405 nm laser that passes through a non-linear crystal. Through the process of spontaneous parametric down-conversion (SPDC), some of the 405 nm photons are converted into pairs of infrared photons, each with half the energy of the photon coming from the laser. Due to the non-linear process, this pair of photons is entangled in both energy and momentum (and possibly also polarization depending on which crystal is used). Thus, when one of the photons in a pair is detected, the other must also exist at a time and location determined by the first. The photons can be measured together in what is called a coincidence measurement. The states of the entangled photons can be measured directly to demonstrate a violation of Bell's inequality. Alternatively, one of the photons can indicate the existence of the other, and the pair can be used as a heralded single-photon source to investigate the wave-like and particle-like nature of single photons. A schematic of a typical experimental setup for the single-photon interferometer experiment is shown in Fig.~\ref{fig:experimentalSetup}(a). Although these experiments are often considered primarily optics experiments, there are electronics and programming components as well. A photo of an experimental realization of the single-photon interferometer is shown in Fig.~\ref{fig:experimentalSetup}(b) to illustrate the level of complexity of the experiment.

For the past two decades, there has been considerable work describing various variations of the single-photon experiments in a manner accessible to undergraduate students. Early papers on the use of these experiments in instructional labs describe the experiments undergraduate students can perform to demonstrate violations of Bell's inequality \cite{dehlinger2002entangled, carlson2006quantum}, single-photon interference, quantum eraser, two-photon interference, and the way a single-photon passing through a beamsplitter can  be detected at only one output (what is often informally called ``existence of a photon,'' a phrase we use throughout this paper) \cite{holbrow2002photon,thorn2004observing,galvez2005interference,gogo2005comparing,pysher2005nonlocal}. These experiments were later extended in a variety of ways to include other methods of violating local realism \cite{gadway2008bell,guzman2015contrasting,beck2016witnessing}, existence of a photon experiments characterizing different kinds of light \cite{scholz2018undergraduate}, Hong-Ou-Mandel interference \cite{carvioto2012hong}, delayed choice \cite{ashby2016delayed,castrillon2019time}, quantum key distribution \cite{bista2021demonstration}, quantum randomness \cite{bronner2009demonstrating}, and single-photon interference through a double slit \cite{lukishova2022fifteen}. Some of these experiments have even been converted into remote versions \cite{galvez2021remote}. Other papers have focused on providing clear explanations of the single-photon experiments for different levels of students, such as second-year undergraduate students \cite{pearson2010hands} and high-school students \cite{scholz2020classical}. There have even been some college-level textbooks written with a large emphasis on these lab experiments \cite{beck2012quantum, waseem2020quantum}. 

Although almost all of the previous work has focused on pedagogical explanations and procedures for using the single-photon experiments, a few instructors have additionally performed investigations into student learning with the experiments in their courses. Lukishova showed that after working with the experiments, students in courses at various levels responded correctly to many conceptual questions and were more interested in a career in quantum optics and quantum information than prior to the lab experience \cite{lukishova2022fifteen}.  Galvez showed that students found the experiments ``striking'' and gave high ratings for their self-reported learning \cite{galvez2010qubit}. Pearson and Jackson found that students were enthusiastic about the experiments and that they were motivated to better understand the theory \cite{pearson2010hands}. However, to-date, there has been no large-scale assessment of the efficacy of these experiments. Many instructors at different institutions are currently interested in having their students work with these experiments \cite{borish2022seeing}, but what do students learn from working with them? Although this paper will not directly answer that question, it will provide the necessary background to be able to do so.

\subsection{Student conceptual learning about quantum optics experiments}

One of the reasons the single-photon experiments are incorporated into undergraduate classes is to improve students' conceptual understanding of quantum mechanics. Quantum mechanics is often thought of as particularly mathematical \cite{johansson2018shut, johansson2018undergraduate, corsiglia2020characterizing}, counter-intuitive \cite{corsiglia2023intuition, singh2009cognitive}, and unrelated to the real world \cite{hoehn2017investigating, dreyfus2019splits}, so one possible way to help students understand these difficult concepts is to explicitly discuss foundational experiments. In order to teach about concepts such as particle-wave duality and entanglement, some instructors have incorporated discussions into their courses about the single-photon experiments, using them either as thought experiments or by discussing the first realizations of the experiments \cite{dhand2018understanding}. Recently, a conceptual assessment about students' understanding of quantum optics through the lens of these experiments has been created \cite{bitzenbauer2022assessing}. 

Another method of improving students' understanding of many of the concepts covered in the single-photon experiments is through the use of simulations. One set of simulations, the Quantum Interactive Learning Tutorials (QuILTs), has been shown to be effective at teaching upper-level undergraduate and graduate students about particle-wave duality, violations of local realism, and other quantum concepts  \cite{marshman2016interactive,marshman2017investigating,maries2017effectiveness}. The Quantum Mechanics Visualization Project (QuVis) is another set of simulations that includes many of the single-photon experiments and has improved learning for students at both the introductory and advanced undergraduate levels \cite{kohnle2013new, kohnle2015enhancing}. Other simulations of these experiments have been created specifically for high-school students \cite{malgieri2016learning}. There are additionally virtual optics labs, where instead of having only specific modules that teach students about individual topics, students may design their own optics experiments using the same components as in the single-photon experiments \cite{la2021virtual,migdal2022visualizing}.

Other instructors use photos or videos of real experiments and real experimental data to help teach students foundational concepts in quantum mechanics. For example, a video of a real experimental apparatus using a down-converting crystal to demonstrate double slit interference with single-photons has been made publicly available for educational purposes \cite{aspden2016video}. Some photos of real versions of the single-photon experiments have also been converted into interactive screen experiments (multimedia representations with which the students can interact to manipulate some aspects of the experiment) \cite{bronner2009interactive}. These interactive screen experiments are particularly useful for students who do not have access to a quantum optics lab. They have been used in German high schools to decrease the influence of classical concepts on students' conceptions of quantum physics \cite{bitzenbauer2021effect}. The extent to which these representations of experiments can be used to replace real experiments has not yet been studied.

\subsection{Other learning goals for working with experiments}

Improving conceptual learning is not the only reason instructors use experiments in undergraduate courses. Lab courses have a variety of other goals \cite{kozminski2014aapt} including helping students learn lab skills, develop expert-like views about the nature of science \cite{wilcox2017developing}, improve communication and collaboration skills \cite{hoehn2020framework,moskovitz2011inquiry}, and build a physics identity \cite{irving2014conditions}. Prior research focusing mostly on introductory labs has shown that instead of reinforcing physics concepts, labs may be more effective at teaching students expert-like experimentation practices, attitudes, and beliefs, and critical thinking skills \cite{holmes2017value,wilcox2017developing,smith2020direct,walsh2022skills}. BFY lab courses in particular tend to be more likely to have learning goals related to lab skills than conceptual learning \cite{holmes2020investigating}. Lab skills is a broad term that may encompass modeling \cite{zwickl2015model,dounas2018modelling}, troubleshooting \cite{dounas2016investigating}, designing experiments, and a large array of technical skills, such as using common measurement equipment and acquiring data with a computer \cite{kozminski2014aapt,zwickl2013process}. In this work, we investigate which of these many goals instructors hope their students will accomplish by working with the single-photon experiments.

\section{Methodology}

In order to understand the full range of ways the single-photon experiments are currently being used, we partnered with ALPhA to engage instructors who use these experiments in their courses.  We first circulated a survey and then performed follow-up interviews of all interested participants. The details and limitations of these two methods are explained in this section.

\subsection{Survey}

The survey was designed around prior work and discussions with instructors who use the single photon experiments. Prior to survey creation, we familiarized ourselves with the published papers about these experiments in educational settings, as well as prior unpublished studies of instructor use and student conceptual learning with these experiments. We used these data as well as informal conversations with instructors to create the questions and answer options on the survey. Once the survey was almost finalized, we discussed it with an instructor who has developed many of the single-photon experiments as used in undergraduate courses. This allowed us to ensure the terminology and questions would be relevant for instructors. That conversation resulted in a few minor changes to the wording and answer options on the survey.  

Each instructor filled out a survey that had several sections that aimed to gather information from multiple courses and experiments. The survey began asking for information about the courses (up to a maximum of two) in which the instructors used the single-photon experiments. This included the course name, the level and major of the intended students, and the option to link to, or upload, a course syllabus. For each course, the survey then asked the instructors which experiments they utilized, using the descriptions shown in Table~\ref{tab:exptNames}. Different instructors implement these experiments in different ways and may combine some of the experiments that we have described separately, but we tried to use the most common names and delineations between experiments. The survey then asked the instructors which actions the students performed, giving instructors the option to select different actions for each experiment. The last part of the survey asked about the importance of a variety of learning goals, letting instructors select \textit{not important}, \textit{somewhat important}, or \textit{very important} for each goal for each experiment. When instructors marked that they used the single photon experiments in a second course, the questions about the experiments, actions, and goals were repeated for the second course. Almost all of the questions were closed response with some open-response options where instructors could add an experiment, action, or goal not included in the closed response options.

\begin{table*}[htbp]
\centering
\caption{Description of individual experiments that can be part of the sequence of the single-photon experiments.}
\label{tab:exptNames}
\begin{tabular}{>{\hangindent=1.5em}>{\raggedright}p{4.4cm} p{10cm} c} 
    \hline
    Experiment name & Description & Sample references \\
    \hline \hline
    Spontaneous parametric down-conversion (SPDC) & Aligning the down-conversion crystal and optics, measuring coincidence counts. & \cite{beck2012quantum} \\
    Existence of a photon & Setting up an extra beamsplitter in front of one of the down-converted paths and measuring correlations, possibly to show there are no 3-way coincidences or to measure the Hanbury Brown-Twiss effect. (Also called the Grangier experiment.) & \cite{galvez2005interference,beck2012quantum} \\
    Photon Stern-Gerlach & Sending one of the photons through a polarizing beamsplitter, detecting both outputs of the beamsplitter, and varying other polarization optics in the path to see where the photon gets detected. (Also called quantum state measurement.)& \cite{beck2012quantum} \\
    Single-photon interference & Seeing interference fringes when one of the down-converted photons is sent through an interferometer; possibly also seeing interference fringes appear and disappear as the length of one arm of the interferometer and/or the spectral filters are changed. & \cite{galvez2005interference,beck2012quantum} \\
    Quantum eraser & Putting half-waveplates or polarizers into the interferometer arms to make them distinguishable and then a polarizer after the interferometer to erase the distinguishability thereby eliminating and reviving the interference fringes. & \cite{galvez2005interference, beck2012quantum} \\
    Delayed choice & Adding parts to the setup so that a choice to measure either the particle or the wave nature of the light is made after the light in the interferometer has been detected. & \cite{ashby2016delayed,castrillon2019time} \\
    Bell's inequality & Measuring polarization states of entangled pairs of photons to apply the Bell inequality to prove that local realism does not hold in quantum mechanics. & \cite{dehlinger2002entangled,beck2012quantum} \\
    Hong-Ou-Mandel interference & Sending two photons at a time into a beam splitter or interferometer to measure the HOM dip or other kinds of two-photon interference. & \cite{carvioto2012hong} \\
    Quantum key distribution & Implementing a quantum key distribution protocol, such as BB84. & \cite{bista2021demonstration}\\
\end{tabular}
\end{table*}

In order to recruit instructors for our study, we emailed the survey to all the instructors who had mentored or participated in one of the ALPhA Immersion workshops between 2010 and 2021, as well as instructors who had purchased single-photon detectors at a discounted price through ALPhA. In total, we emailed 170 instructors. Additionally, we posted the survey on the ALPhA Slack channel. We received complete responses from 28 instructors, one of whom submitted two responses in order to provide data about three different courses. These instructors were working at 27 unique institutions, which are described in the upper half of Table~\ref{tab:demographics}. 

\setlength{\tabcolsep}{1.8pt}
\begin{table}
\centering
\caption{Types of institutions and self-reported demographics of interview participants. Two of the surveyed instructors worked at the same institution but reported about different courses, so their institution is counted twice. We did not collect demographic information for the surveyed instructors.}
\label{tab:demographics}
\begin{tabular}{>{\quad}lcc}
    \hline
    \hline
    & \makecell{Survey\\(N=28)} & \makecell{Interviews\\(N=14)}\\
    \Xhline{2\arrayrulewidth}
    \rowgroup{Institution} \\
    Four-year college & 13 & 7 \\
    Master's degree granting & 5 & 4 \\
    PhD granting & 10 & 3 \\
    Hispanic-serving institution & 3 & 1 \\
    U.S. & 24 & 14 \\
    International & 4 & 0 \\
    \hline
    \rowgroup{Instructor}\\
    Man & - &  12 \\
    Woman & - & 2 \\
    White, not Hispanic & - & 12 \\
    White, Hispanic & - & 2 \\
   \hline
    \hline
\end{tabular}
\end{table}

The data in the survey were divided up by instructor, course, and specific experiment, so we combined the data in different ways for our analysis. We wanted to investigate how much variation there was across course types and different experiments, while still presenting overall trends. We choose to look at the data by course instead of by instructor because many instructors use the experiments in distinct ways in different courses. We classified each course's type based on the course name and uploaded syllabus.  When combining experiments for a single course, we collapsed the data differently for actions taken and learning goals. For actions taken, we determined that students in a course had performed an action if they had performed that action while working with at least one of the experiments. For the goals, we chose to present the highest level of importance for a given goal across all experiments in a course.

\subsection{Interviews}

In order to better understand instructor responses to the survey, we performed follow-up semi-structured interviews of 14 instructors. We invited for an interview all of the US instructors who completed the survey and agreed to be contacted for future research opportunities. Of those, 14 chose to participate. Demographic data about the interviewed instructors is shown in the lower half of Table~\ref{tab:demographics}.
The interviews occurred over Zoom and ranged from 49 to 69 minutes. They contained sets of questions about the course context, ideas related to the notion of seeing quantum mechanics, and other goals the instructors had for the experiments, including student excitement and lab skills. The questions related to seeing quantum mechanics have been preliminary discussed in prior work \cite{borish2022seeing} and are the subject of future work, so here we focus on the other sections of the interviews. Sample relevant questions include:
\begin{itemize}
\setlength\itemsep{0em}
    \item What background material related to the concepts of the single-photon experiments do you cover in the non-lab portion of your course (for example, in lectures, pre-labs, etc.)?
    \item In the survey, you responded that the goal of ``Be excited about physics and/or motivated to learn more'' was [their survey response].  What specifically is your goal for the students related to this?
    \item What experimental skills do you want students to gain from working with these experiments?
\end{itemize}
To analyze the interviews, we performed a thematic coding analysis of the transcripts using both a priori codes coming from the closed response survey options and emergent codes discussed by the research team.

\subsection{Limitations}

One limitation of this work is that our sample of instructors may not be fully representative of all instructors using similar experiments. The only methods we had of reaching instructors using the single-photon experiments were through ALPhA, so we may not have had a chance to hear from instructors who developed similar experiments on their own or who utilized the experimental set-up in an entirely different way. Additionally, the majority of the instructors we interviewed were very enthusiastic about these experiments and had invested a lot of time to get them working and implemented in their courses. Instructors who are more eager to participate in research about these experiments may be more invested in the experiments than other instructors. When we sent out the survey, we received email responses from some instructors saying they could not fill out the survey because they were not able to get the experiments up and running due to a variety of barriers, including lack of money, suitable space, and time. It is therefore important to keep in mind that the data presented in this paper is coming from only the instructors who were able to successfully fund and set up the experiments and who had been associated with ALPhA to some degree. Additionally, we sub-divide the data we have to see themes by both course type and experiment. This leads to categories with small sample sizes that do not have much statistical power. Nonetheless, this work presents many concrete, although not exhaustive, possibilities for how the single-photon experiments can be used, and the distinctions by course type or experiment give an idea of the kinds of trends that may exist in a larger sample.

\section{Incorporation of experiments into courses} \label{sec:howUsed}

Our first set of results answers RQ1 and covers how the single-photon experiments are currently implemented in courses. We present survey data about the courses these experiments are used in, how many and which of the sequence of experiments are most commonly used in different kinds of courses, and what actions the students perform while working on the experiments. Each of the survey respondents described the use and goals of these experiments for between one and three distinct courses, leading to a total of 37 courses in this dataset. We supplement the survey results with details coming from the interviews to better understand some of the survey responses and the larger context in which these experiments are used.

\subsection{Courses} \label{sec:courses}

Although there are a variety of course types in which the single-photon experiments are used, they fall into two main categories. We classified eight of the courses in the survey data as quantum mechanics courses primarily for junior and senior physics majors. These are lecture courses with lab components (at least half of the course is non-experimental), where the entire focus of the course is on quantum topics. This includes courses such as the first or second semester of upper-level quantum mechanics or a quantum optics course. We categorized another 25 of the courses as BFY lab courses. These are lab courses for students majoring in physics, applied physics, or engineering physics who are beyond the first year of study in their major. These courses may include a short lecture or discussion component that is focused on the concepts or skills needed to complete the lab experiments. Almost all of these courses are intended for juniors and/or seniors with two also including graduate students and one of the courses being only for sophomores. Many of the courses were named with some variation on ``Advanced Lab,'' which is a common type of course at institutions in the United States.

There were four additional kinds of courses reported on the survey, namely an introductory lab course, an introductory non-lab course, a senior capstone course, and an undergraduate thesis. Since there were so few courses of these other types, we focus on the quantum and BFY lab courses when we divide the results by course type. Even though we are able to classify the courses in primarily two broad categories, instructors use the experiments in diverse ways within each category, as described further below.

\subsection{Prevalence of experiments}

\begin{figure*}[htbp]
  \centering
  \includegraphics[width=\textwidth]{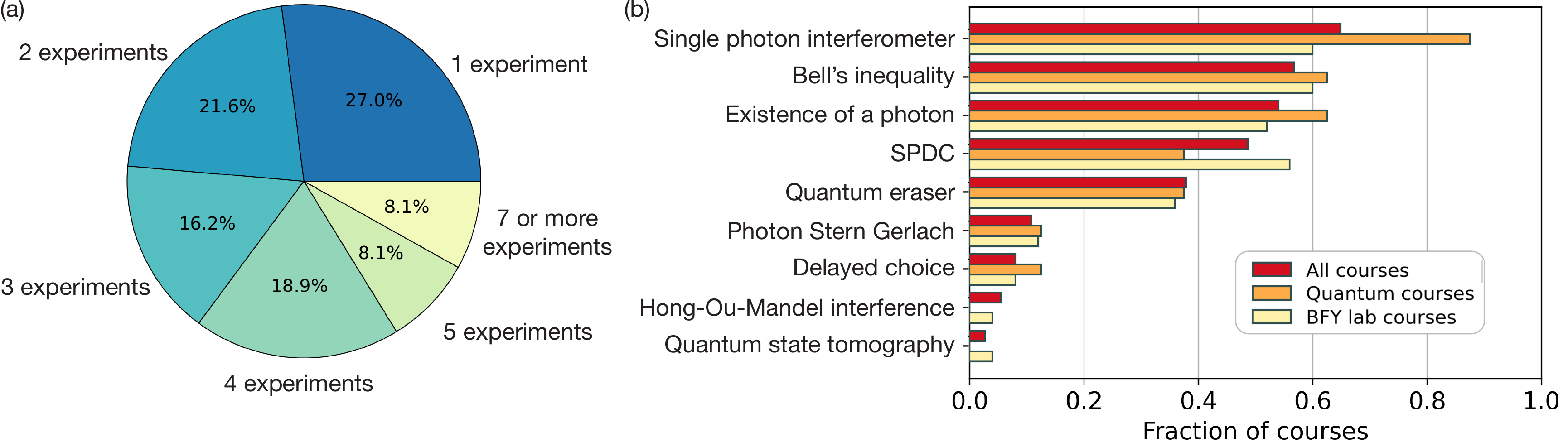}
\caption{(a) Number of experiments used per course (N = 37). (b) Fraction of all courses (red, N=37), quantum courses (orange, N=8), and BFY lab courses (yellow, N=25) in which each of the different experiments were used, sorted by fraction of all courses.} 
\label{fig:whichExperiments}
\end{figure*}

The single-photon experiments consist of a set of experiments that may be performed on one experimental apparatus either as a sequence or individually. Some instructors choose to perform the entire sequence (or most of it) while others focus on just one, or a sub-set, of the experiments that best match their learning goals and the time available in the course. In order to know which experiments would be most useful to study student learning in the future, we investigated which experiments instructors used most frequently. Figure \ref{fig:whichExperiments}(a) shows the number of experiments used in each of the surveyed courses. Almost half of the courses used only one or two experiments, whereas the others incorporated different amounts, even up to seven experiments or more. 

Figure~\ref{fig:whichExperiments}(b) shows the prevalence of experiment use in all of the surveyed courses, in the quantum courses, and in the BFY lab courses. The most commonly used experiment is the single-photon interferometer, which is one of the experiments often taught in the ALPhA Immersion workshops. The five most common experiments (single-photon interferometer, Bell's inequality, existence of a photon, SPDC, and quantum eraser) were each used in 14 or more courses, whereas the others (those displayed, as well as quantum key distribution, which was not performed in any of the courses) were used in less than five courses each. The single-photon interferometer was especially popular in quantum courses, and the SPDC experiment was used by students in BFY lab courses more than students in quantum courses. The SPDC experiment includes the initial set up of the apparatus, so it may be more aligned with the goals of BFY lab courses. Although there are slight differences in experiment use between quantum courses and BFY lab courses, the five most common experiments overall are the five most common experiments in both of those course types as well. 

Although there are clear trends in the survey data about which experiments are used most frequently, the order of, number of, and time allocated for the experiments varied by course, as described in the instructor interviews. The quantum courses often implemented several of the experiments as one or two day labs. Some of the BFY lab courses implemented these experiments in a similar way, whereas others used them as multi-week projects ranging from three to 12 weeks. In both course contexts, many courses focused on photon particle-wave duality (the existence of a photon and single-photon interferometer experiments) or Bell's inequality. Additional details about the timing and sequencing of the experiments in different course types can be found in Appendix~\ref{sec:implementationCourseType}.

\subsection{Actions performed} \label{sec:tasks}

\begin{figure*}[htbp]
  \centering
  \includegraphics[width=\textwidth]{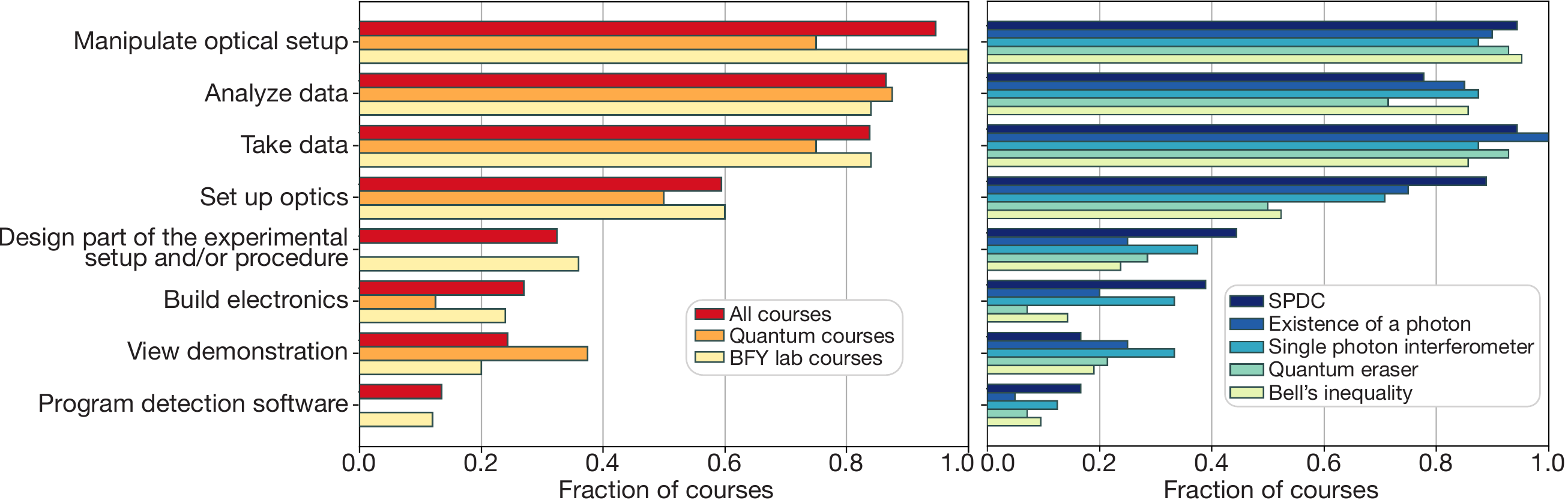}
\caption{Actions students performed. On the left is shown the fraction of all (red, N=37), quantum (orange, N=8), and BFY lab (yellow, N=25) courses in which students performed the specific action in at least one of the experiments. 
On the right, with the same vertical axis, is shown the fraction of courses that perform the experiments SPDC (N = 18, darkest blue), existence of a photon (N=20), single-photon interferometer (N = 24), quantum eraser (N = 14), and Bell's inequality (N = 21, lightest green) in which the students performed the specific action.}
\label{fig:howUsed}
\end{figure*}

There was also variation in the actions the students took while working with the single-photon experiments. Figure~\ref{fig:howUsed} shows the common actions by both course type and experiment in the surveyed courses. Overall, the four most common actions consist of setting up and manipulating optics and taking and analyzing data. Note that it is possible that in some courses, the experiments were not set up enough for the students to perform certain steps. For example, one instructor explained on the survey that the students were still getting the experiment working so they were not yet able to analyze data even though the instructor viewed it as an important step. Instructors assigned other activities, which they added as write-in options, including reading journal articles, communicating results, responding to questions in lab write-ups, and deriving theoretical components of the experiment.

The left panel of Fig.~\ref{fig:howUsed} shows that there were only small differences in student actions based on course type. One distinction was that students in quantum courses were more likely to view demonstrations instructors conducted than students in BFY lab courses, and students in BFY lab courses were more likely to manipulate the optics. Additionally, none of the quantum courses surveyed had students design part of the experimental setup or procedure or program any of the detection software. These were not commonly performed actions in BFY lab courses either, but they were performed in some. 

There were also small differences between actions students performed while working on the different experiments, as shown in the right panel of Fig.~\ref{fig:howUsed}. One experiment that stood out from the other experiments is SPDC. Instructors were more likely to have had their students set up optics with that experiment than with any of the others. The SPDC experiment entails setting up the apparatus and obtaining coincidence counts, so this result is not surprising. Similarly, students were more likely to design part of the experiment, build electronics, or program the detection software while working on the SPDC experiment. The single-photon interferometer was the experiment most used for demonstrations.

Although almost all implementations of these experiments involved manipulating or setting up optics in some capacity, there was a large spectrum of the amount of optical manipulation the students performed, as discussed by instructors in the interviews. This ranged from the students only tweaking already set-up optics (e.g., controlling a piezo mounted on a mirror and a polarizer) to performing the entire optical alignment themselves. Many of the courses were somewhere in between. For example, the instructors roughly aligned the setups and then allowed the students to optimize the alignment or had some parts (e.g., the lasers) already mounted on the optical table but allowed the students to add the rest. The decision of how much optical alignment the students performed was often seen as a compromise between helping the students gain experience with optical alignment and performing the experiment in the allotted time. Strategies instructors took to find the right balance are discussed in Sec.~\ref{sec:challenges_alignment}.

In contrast to manipulating and setting up optics, taking and analyzing data were steps that took a relatively short amount of time for the students once the experiments were set up. Most of the coding was already done for the students, with code obtained from another instructor or prior students; however the amount of data analysis performed depended on the course. Instructor 1\footnote{Due to the small number of instructors working with the single-photon experiments and in order to ensure anonymity of the interview participants, we have chosen to remove gendered pronouns and avoid pseudonyms with any gender, racial, or ethnic connotations.}, when asked what piece of equipment they wanted students to gain experience working with, discussed how they changed the way students performed data analysis as they used this experiment over time:
\begin{quote} 
    \textit{``It used to be that the LabVIEW code that I have would just kind of spit out the answer... And I stopped doing that, and now they just get the raw data, and then they need to analyze it. And I think that that's now actually... working a lot better... If you want to be an experimental physicist, you need to know how to analyze data. And you can't analyze the data unless you understand the experiment that it came from...''}
\end{quote}
They went on to say that this helped students both learn data analysis techniques and better conceptualize the experiment.

\section{Goals for the experiments} \label{sec:goals}

\begin{figure*}[htbp]
  \centering
  \includegraphics[width=\textwidth]{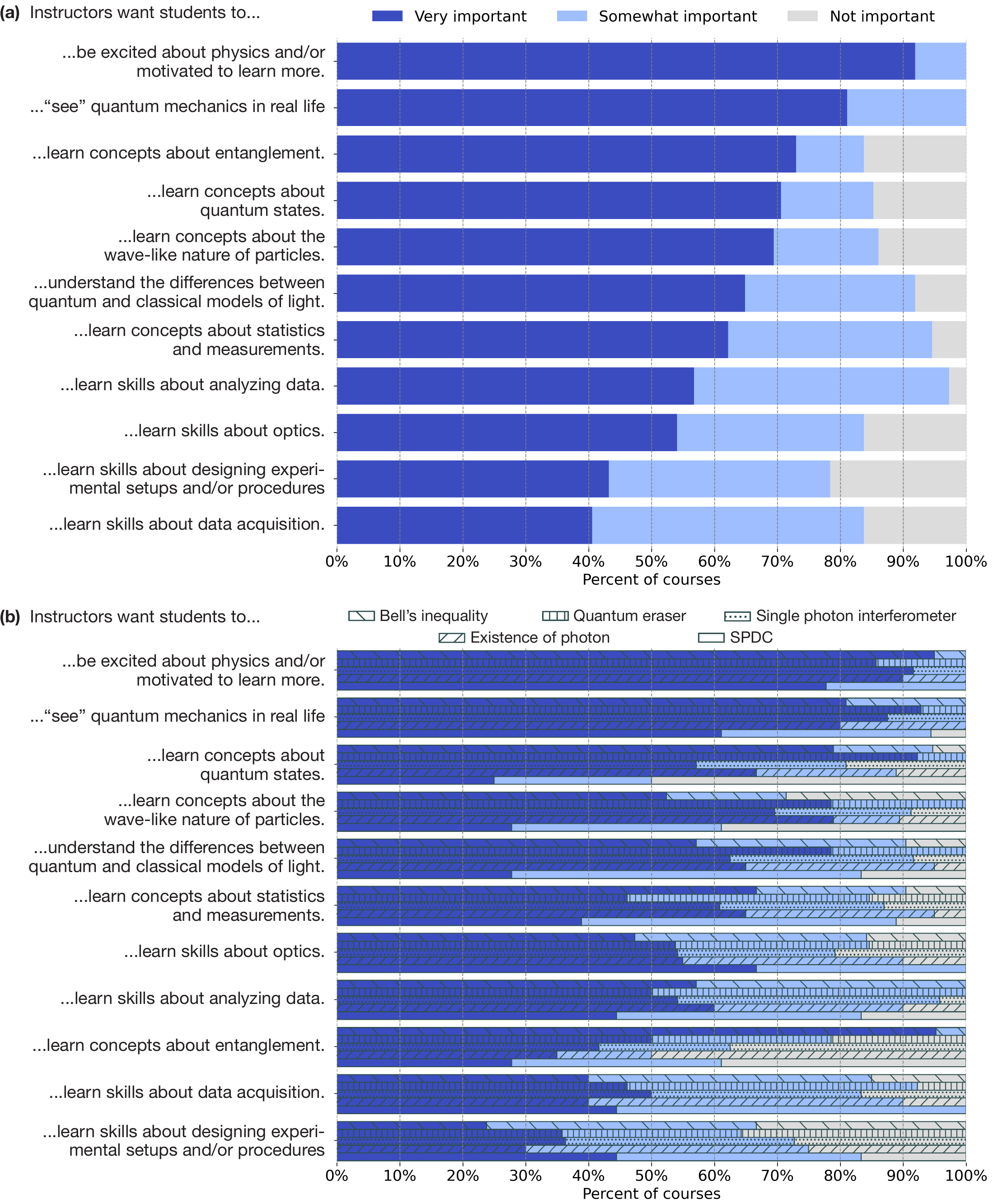}
\caption{(a) Percent of courses in which the instructor marked very important (dark blue), somewhat important (light blue), and not important (gray) as the highest level of importance for the given goal over all experiments used in their course. Due to a few missing responses, the number of courses N = 34 for the goal of learning concepts about quantum states, N = 36 for the goal of learning concepts about the wave-like nature of particles, and N = 37 for all other goals. (b) Percent of courses using each experiment in which the instructor ranked each goal as very important (dark blue), somewhat important (light blue), and not important (gray) for the five most common experiments (distinguished by hatching). The goals are sorted by the average percentage of very important rankings over the five experiments.}
\label{fig:goals}
\end{figure*}

Our second set of results answers RQ2 and demonstrates that instructors have a wide variety of learning goals for using the single-photon experiments. In the survey, we asked instructors to rank potential learning goals related to student affect, conceptual learning, and learning of lab skills for each experiment. Overall, instructors ranked the majority of the goals as somewhat or very important, demonstrating that they value many aspects of the experiments. Figure~\ref{fig:goals}(a) shows the highest level of importance of each goal across all experiments for the courses in our dataset. By that, we mean that a course was assigned the response of very important for a given goal if the instructor marked that goal as very important for at least one of the experiments. Similarly, if the instructor did not mark any experiments as very important for a given goal, and they marked the goal as somewhat important for at least one of the experiments, then it was assigned a response of somewhat important. A goal assigned not important means that it was marked as not important for all experiments. Additional goals related to professional skills and epistemological views emerged in the interviews. Here we briefly describe instructors' goals for these experiments with additional details provided in Appendix~\ref{sec:detailedGoals}.

The two goals with the most rankings of very important across all courses were ``be excited about physics and/or motivated to learn more'' and ```see' quantum mechanics in real life.'' All of the instructors thought these goals were at least somewhat important, so we focused on these goals in the follow-up interviews. From the interviews, we know that many instructors hoped these experiments would motivate students to spend time on their coursework and pursue physics in the future. When asked what their goal related to student excitement and motivation was, Instructor 2 discussed the way they wanted students to experience the authentic excitement researchers have:
\begin{quote} 
    \textit{``I try to pick experiments that I think will spark some curiosity and spark some of that authentic excitement you get in grad school when you're doing an experiment, you see something for the first time, and it makes sense to you. I guess I'm trying to engender that excitement and that willingness to recognize that you don't know something, but you can figure out the answer.''}
\end{quote}
Excitement and motivation was one of many themes tied in with the idea of seeing quantum mechanics, which is described in Ref.~\cite{borish2022seeing} and will be detailed further in future work. 

Many of the other goals ranked as very important by the majority of the instructors on the survey were conceptual. Learning about entanglement was the conceptual goal indicated as very important by the most instructors, although this is mostly due to the Bell's inequality experiment (see Fig.~\ref{fig:goals}(b)). The goal of learning about concepts related to statistics and measurement was the conceptual goal indicated as either somewhat or very important by instructors of the most courses. Other conceptual goals include learning about quantum states, particle-wave duality, and quantum versus classical models of light. Some instructors additionally wanted their students to make the connection between the concepts and the mathematical formalism while working with the single-photon experiments. Instructor 3 mentioned that one of the unique aspects of these experiments is that undergraduate students can do the full theoretical derivation alongside the experiments:
\begin{quote} 
    \textit{
    ``[In] these experiments, you can write down the wave function essentially and follow it through. So I think that's maybe what's a little bit different than a lot of the other experiments...you have to use the quantum formalism, the math, a little bit to solve, to analyze the data or to explain the data... I just can't think of other experiments where you really do that at this level.''}
\end{quote}

Although fewer instructors ranked them as very important compared with the conceptual learning goals, all of the lab skill goals on the survey were marked as somewhat or very important by the majority of the instructors. These included skills about data analysis, optics, experimental design, and data acquisition. Data analysis was ranked as the most important skill, with many instructors focusing on students' understanding and analyzing coincidence measurements and knowing how to tell if measurements are correlated with each other. When asked what skills they hoped students would gain from working with the single-photon experiments, Instructor 4 discussed data interpretation: 
\begin{quote} 
    \textit{``I would definitely love students to be able to...interpret what they're seeing on the screen in terms of the counts and the coincidence counts that they're seeing... be able to translate what's happening on the table into the data that they're seeing and then be able to interpret the uncertainties or the confidence and making the claims that they're making.''}
\end{quote}

Since the single-photon experiments can be implemented as multiweek projects, instructors often hope students will gain some of the professional skills needed in research as well. These skills came up unprompted in the interviews and included intrapersonal skills, such as independent decision-making and persistence, and interpersonal skills, such as teamwork and leadership. For example, when asked about any additional goals they had for their students, Instructor 2 discussed grit and perseverance:
\begin{quote} 
\textit{``And just trying to teach them the importance of grit, the importance of persevering in the face of adversity... A lot of the students have a very low tolerance for frustration... I'm just trying to instill in them the sense of you know, this is what you're gonna have to deal with. You're going to be thrown into a situation where you're going to get frustrated.''}
\end{quote}
Being able to communicate scientific results was another goal mentioned by many instructors, with the focus ranging from short research-group-meeting-type presentations to formal written reports. These professional goals are not unique to the single-photon experiments, but the complexity of the experiments provides the opportunity for students to practice them. 

The other goal that emerged in the interviews was helping students recognize that physics is an experimental science. Instructors wanted students to realize that what they learn in their theoretical quantum mechanics courses describes the world, as seen in the lab. When asked about the importance of seeing quantum mechanics, Instructor 4 more broadly discussed the experiment-theory connection:
\begin{quote} 
    \textit{``You have lab courses and you have theory courses, and they seem to be separate from each other. And this idea that physics is an experimental science, and certainly that people do all sorts of amazing work purely in theory, but that they have to... meet up somewhere. And that no matter what theoretical physics course you're taking, there is a corresponding experiment that you can do.''}
\end{quote}
Instructors additionally wanted students to understand that questions about the world are answered through experimentation. When asked about additional goals for their students, Instructor 5 talked about the goal of ``giving primacy to data'': students learning how to draw unbiased conclusions from data. 

Although we focus our discussion on the possible goals across all courses, there may be slight differences in goals between course types and populations of students. In the interviews, some instructors differentiated between their goals for quantum courses and BFY lab courses. When asked how the students worked on the experiments in their advanced lab course, Instructor 1 said:
\begin{quote} 
    \textit{``In the quantum optics class... I actually want to make sure that they get through [the experiment] and see what I want them to see. And that's more the goal. Whereas in the advanced lab, the goal is more playing around with stuff and learning experimental techniques. So they go a little bit slower in there...''}
\end{quote}
Some instructors also mentioned differences in their goals for students based on their intended career path: engineers may be more interested in materials, optics, or sensors for detecting low levels of light, whereas physicists may be more interested in learning about quantum concepts. 

The survey data additionally shows some differences in goals for the different experiments. Figure~\ref{fig:goals}(b) shows the percentage of all courses using a given experiment in which the instructors gave ratings of very important, somewhat important, and not important to the goals for the five most common experiments. The SPDC experiment is notable in that the conceptual goals are ranked as very important in a smaller percentage of courses, compared with the other experiments. Additionally, the skills goals of optics and experimental design are ranked as more important for SPDC than for the other experiments. Another difference between experiments is for the goal of conceptual learning about entanglement. This goal is ranked as very important in almost all of the courses that use the Bell's inequality experiment, but it is only very important for half or fewer of the courses for the other experiments. All of the experiments use pairs of entangled photons, but instructors care most about students learning that concept while performing the Bell's inequality experiment.

\section{Overcoming experiment challenges}

Throughout the interviews, we observed themes of common challenges instructors faced while using the single photon experiments. Two challenges that stood out as particularly salient were the difficulty instructors had with:
\begin{enumerate}[label=\Alph*.]
    \item Providing students the requisite knowledge and skills to understand and perform the experiments
    \item Helping students learn technical skills (e.g., optical alignment) while also completing the experiments in fixed amounts of time.
\end{enumerate}
Although we did not design the survey or the interview protocol to investigate instructors' challenges--- our goal, instead, was to discover what instructors wanted students to learn so we could then focus on investigating student learning gains---, we realized this is an important topic for instructors. In this section, we therefore describe various strategies instructors took to overcome these challenges, in the hope that sharing these lessons will benefit other instructors and motivate researchers to examine these aspects in future work.

\subsection{Student preparation}

In order for students to understand and use the single-photon experiments, they need some knowledge of both quantum concepts and how the experimental apparatus works. In some courses, students had already gained this specialized knowledge either through the course (especially for quantum courses) or through prior required courses. However, that was not always the case, especially when the single-photon experiments were used as stand-alone experiments in BFY lab courses. Even for the courses that did require students to have previously taken quantum mechanics or modern physics courses, not all pre-requisite courses covered the quantum optics concepts and math necessary to understand the single-photon experiments.

The instructors interviewed used a variety of strategies to provide their students the necessary concepts and mathematical formalism to work with these experiments. For the quantum courses, that material was covered in the lecture portion of the courses. The lab courses often provided various resources to introduce the students to both the theory and the apparatus. These included videos made by the instructors, readings about the experiments (e.g., Refs.~\cite{pearson2010hands, galvez2005interference, beck2012quantum}), readings about statistics (e.g., discussing normal and Poisson distributions), homework assignments, papers from prior students who had worked on the same experiment, and lab manuals. Many of these courses had the expectation that students would learn on their own from the readings, and the instructor would offer additional one-on-one help as needed. 

In additional to conceptual knowledge, there are many lab skills needed to perform the single-photon experiments, which are often also new to the students. Although one of the BFY lab courses required all students to take a prior optics or electronics lab course, most of courses did not have specific requirements about optics skills from prior courses. Although it was not always feasible due to the course structure, some of the BFY lab courses required students to work on certain other experiments before progressing to the single-photon experiments. For example, some courses required students to learn how to use classical optical elements (e.g., polarizers and wave plates) or to perform classical versions of experiments analogous to the single-photon experiments (such as an interferometer with a visible laser). This was used to teach students optics skills, as well as to later compare the classical and quantum versions (such as for the measurement of the degree of second-order coherence in the existence of a photon experiment). Another course had students work on an unrelated nuclear lab prior to working with the single-photon experiments, so the students could learn about coincidence counts. 

\subsection{Trade-off between practicing optical alignment and obtaining results} \label{sec:challenges_alignment}

When students set up the apparatus entirely from scratch, the single-photon experiments take more time to perform than students have during a course. Instructors utilized a variety of strategies to ensure the students were able to obtain data and complete the experiments even with the difficult and time-consuming optical alignment. For shorter guided labs, some instructors chose the order of the experiments such that students built on prior weeks' work, so by the end of the class the students could perform complicated experiments in less time while understanding all the steps. In one class, one of the groups of students performed the alignment for each experiment, and all the other groups used that set up to collect data. The group that performed the alignment would change throughout the term, so all students could practice optical alignment at least once. When used as a project where one group of students would work through the entire sequence of single photon experiments before another group, some instructors left the alignment from a prior group of students. This allowed the new group of students to either start the experiment there or take data at the beginning before mis-aligning and re-aligning everything. That strategy caused the students less stress if they later struggled with the alignment. Many instructors stepped in on an as-needed basis either in class or in between lab sessions to improve the students' alignment. 

One instructor found that converting the experiment to be conducted remotely due to the COVID-19 pandemic avoided the problem of one group of students mis-aligning the set up for other groups when many groups of students in the course were performing the experiments at similar times. The remote format made it easier to reset between groups and have the instructor available to help all the groups. Although the students were not able to perform all of the optical alignment remotely, they were able to remotely control some of the optics and see those parts moving through webcams. Additionally, the instructor thought the results of the remote experiment were better than normal, possibly because the students did not rush out at the end of the lab because they were able to choose the time that best suited them. However, converting the single-photon experiments to be remote is not without challenges. Another instructor also created a remote interface during the pandemic, but did not retain it after in-person instruction resumed because it was not-well optimized.

\section{Discussion and conclusions} \label{sec:conclusion}

Through the instructor survey and interviews, we have found that the single-photon experiments are used in a wide variety of ways, for an assortment of learning goals, and are often tied in with other aspects of the courses or broader departmental curriculum. These differences may make these experiments difficult to study, but they show that there is a rich variety of options available for instructors. There are slight differences in which experiments are done, what actions the students themselves perform, and what the goals are based on the course types and the specific experiments; however there are also overall trends spanning most courses and experiments. The single-photon experiments may be uniquely situated to accomplish some of the desired learning goals, but other goals may be easily accomplished with other complex experiments integrated into undergraduate courses. These findings have a variety of implications for both instructors and researchers.

Learning about the varied uses and goals of the single-photon experiments may prove useful for instructors, especially those looking for examples of novel ways to incorporate these experiments into their courses. There are many different goals that can be accomplished with these experiments, and we hope to inspire instructors to think beyond the obvious goals of learning quantum concepts and optics skills to consider emphasizing some of the plethora of other goals that they may also value. Several of the instructors wanted to learn better ways to frame the experiments both at different levels and without the support of a theoretical quantum course, as the experiments are often discussed in that context. Some instructors also struggled to find a way to quickly provide the necessary conceptual and technical knowledge to help students engage with some of the more complicated experiments. Although we cannot yet provide definitive recommendations for best practices, we hope this work will contribute to discussions about the challenges instructors face and illuminate possible solutions currently being implemented in other courses.

For researchers trying to evaluate the efficacy of these experiments, it is important that they choose both specific learning goals to target, as well as a specific context in which to study those goals. There does not seem to be a single way the single-photon experiments are used, as one might expect based on the way many of the instructors who utilize these experiments learned about them through the ALPhA Immersion workshops. The experiments are commonly used both as short, guided labs in quantum and BFY lab courses and as longer projects in BFY labs. The goals for these uses and courses often differ. Although the single-photon experiments are often discussed as being used to teach concepts, instructors use these experiments to additionally teach technical and professional skills, help students develop expert-like views about the nature of science and experimentation, and improve student affect. Additionally, it is important to realize that these are not one-off experiments; they depend on students' prior conceptual knowledge and lab skills, as well as the other experiments performed and resources provided in the course. Researchers will have to carefully decide what student outcomes they want to measure and understand the students' backgrounds before designing the appropriate assessments, keeping in mind that different instructors focus on different goals when initiating a study across many course contexts. 

This work is just the first step in researching the potential benefits the single-photon experiments may have for undergraduate students. We hope that these data can be used to plan out further studies investigating whether or not students are achieving the specific learning goals set out by their instructors and whether the learned concepts and skills are transferable. More work needs to be done studying students to show whether these experiments are worth the time and money needed to implement them. Follow-up studies could investigate which student outcomes are unique to these experiments, and if those could be achieved in less resource-intensive ways, helping to make experimental quantum physics more accessible to a larger number of students.

\begin{acknowledgements}
We thank the instructors who responded to our survey and participated in interviews, and the CU PER group for useful conversations and feedback. This work is supported by NSF Grant PHY 1734006 and NSF QLCI Award OMA 2016244.
\end{acknowledgements}

\appendix

\section{Sequence and timing of experiments by course type} \label{sec:implementationCourseType}

The quantum courses often used the single-photon experiments as a sequence of 1\textendash2 day labs, incorporating between one and five of these experiments in each course. All of the students in the course would perform the entire sequence of experiments in the same order. The sequence of experiments would often include the existence of the photon experiment and the single-photon interferometer (sometimes with the quantum eraser) to demonstrate both particle-like and wave-like aspects of light. Bell's inequality was also commonly included, although in some cases, it was conducted only in a second semester quantum course after other experiments were performed in the first semester quantum course. Other ways these experiments were incorporated into more lecture-based quantum courses was by using them as one-off demonstrations or experimental homework sets.

In BFY lab courses, there were two primary ways the experiments were used: as sequences of short guided labs and as open-ended longer projects. When used as guided labs, the timing and sequences were often similar to what was done in the quantum courses. However, in some BFY lab courses, all of the students would cycle through these guided labs, whereas in others the sequence would be one of many experiment options from which the students could choose. Students usually had a choice of experiment(s) for the open-ended projects as well, so not all students in a course would have the opportunity to work with the single-photon experiments. The projects ranged from 3 \textendash 12 weeks with students working for around 3 \textendash 8 hours a week (either one or two lab sessions), with some courses even providing students around-the-clock lab access. 

Some BFY lab courses focused on one of the single-photon experiments while others incorporated many. When used as guided labs, the courses often included SPDC, existence of a photon, single-photon interferometer, and quantum eraser. Bell's inequality was added at the end to a couple courses as well, in one case only if the students performed the other experiments fast enough. When used as projects, students often performed SPDC on the way to Bell's inequality or some combination of existence of a photon and the single-photon interferometer. A couple of instructors found that the experiments were more successful once they decreased the number of experiments the students implemented to give them more time for each. Some of the courses with guided labs additionally had a project at the end where the students could extend work they had already done or try out a new idea. Examples included performing experiments commonly done in other courses, but not in their course (e.g., performing Bell's inequality in courses that did not otherwise include it), building a different set up for the same experimental goal (e.g., a Michelson interferometer instead of a Mach Zehnder interferometer), and attempting to perform an experiment not commonly done in undergraduate courses (e.g, ghost imaging \cite{pittman1995optical,shapiro2012physics}). Some of the projects consisted of adding on to a part of the experiment built by the prior year's students, so the experimental setup was extended to include more capabilities each year.

The introductory, capstone, and thesis courses used the single-photon experiments in yet other ways. One of the introductory courses used only the quantum eraser experiment as a single three-hour lab session, whereas the other one paired the introductory students with more advanced students for a 12-week-long project. For the semester or year-long capstone and thesis courses, the students developed or built a new experiment or found ways to make the experiments easier to use in other courses.

\section{Experiment goals} \label{sec:detailedGoals}

In this section, we discuss additional details about the instructors' learning goals for using the single photon experiments. We present quotes from the interviews to help understand the goals on the survey and further expound upon the other goals emergent in the interviews.

\subsection{Seeing quantum mechanics} \label{sec:seeingQM}

Prior to our survey, we had many informal conversations with instructors where the idea of ``seeing quantum mechanics'' arose. We therefore included it as a possible learning goal in the survey; however we did not know what instructors meant by it when they ranked it as an important learning goal. During the interviews, instructors expounded upon their definitions of the phrase and the variety of potential learning goals to which it relates. These ideas are briefly described in Ref.~\cite{borish2022seeing} and will be further expounded upon in future work. For completeness, we list the goals here:
\begin{itemize}
    \itemsep0em 
    \item Believe quantum mechanics describes the physical world
    \item Gain familiarity with quantum mechanics
    \item Improve conceptual understanding
    \item Think about the philosophy of quantum mechanics
    \item Generate excitement and motivation
    \item Learn about topics of technological and societal importance
    \item Make quantum more accessible
\end{itemize}

\subsection{Student excitement and motivation}

Student excitement and motivation, the other most highly ranked goal, also elicited from instructors a variety of reasons for why it was important. Being excited about and interested in the single-photon experiments can motivate students in their coursework, helping them to complete the course assignments and spend the time necessary to understand the concepts. Student excitement can also motivate students to pursue physics in the future, whether their interest lies in quantum physics, classical optics and the experimental manipulation of light, or physics more broadly. Additionally, some instructors wanted their students to experience the authentic excitement physicists have when working on new problems. Some instructors mentioned that although student excitement is not an explicit goal of theirs, they believe part of their job as instructors is to help students be interested and excited about the topics they teach.

Instructors thought that students found many different aspects of the single-photon experiments to be exciting. They believed some students were interested in the concepts behind the experiments and their relation to many up-and-coming quantum technologies. The equipment required for the experiment could be fun to use (especially the lasers), be similar to equipment used in cutting-edge research labs, and allow the students to measure something that was invisible to them. The act of performing an experiment could also be exciting for the students whether because students enjoyed building experiments or because these experiments in particular were less of a black box than some other complex undergraduate experiments. Successfully getting the challenging experiments to work could also generate excitement for the students, as described by Instructor 6 when asked about their goal related to excitement and motivation:
\begin{quote} 
    \textit{``Some students... get a lot of satisfaction out of seeing something work. And so when they're like we spent two hours working on making these three degree measurements and then it actually worked, that's like an excitement about doing physics... So I think the student participation, the amount that they put into actually making these experiments happen, leads to more excitement about them, and more frustration.''}
\end{quote}
Additionally, these experiments provided a new experience for the students, allowing them to see something they may have heard about but had never seen before.

\subsection{Technical skills}

The skill that came up most frequently in the interviews was optics, with many instructors wanting their students to learn how to align and build optical setups and work with the related equipment. Some of the specific techniques instructors mentioned included using irises and mirrors to steer lasers, aligning non-visible beams, and aligning light into optical fibers. Some instructors focused on student understanding of basic optical elements (i.e., waveplates and beamsplitters), while others wanted students to have experience with advanced instrumentation such as avalanche photodiodes. One instructor mentioned that knowing how to measure low intensities of light is an industry-relevant skill. The opportunity to learn optics skills was especially appreciated because optics is a topic not focused on in many institutions' physics curricula. How much instructors prioritized optical skills depended on the course, with some variation between quantum and BFY lab courses. 

When discussing data analysis, many instructors discussed learning how to interpret data, especially when dealing with correlated measurements. When asked what skills they hoped students would gain from working with the single-photon experiments, Instructor 7 
said that the primary way they used the experiment was as a ``puzzle of data interpretation.'' Some instructors discussed how the experimental techniques of looking for signals that are coupled together or finding a small effect in a large background can be useful in other contexts. Another instructor instead focused on students learning the difference between Poisson and Gaussian statistics.

For data acquisition, instructors wanted students to learn both hardware and software skills. On the hardware side, many of these experiments use a field-programmable gate array (FPGA) so learning how to use that and associated electronics, as well as sending electronic signals back and forth more generally, was important for some instructors. Others focused on the way students could learn how to interface different pieces of equipment with computers. Another instructor thought it was important for students to recognize that data is acquired with a computer, which makes it much more controllable and reliable than the methods used in introductory physics labs. Additionally students could contemplate how to decide what kind of computer program to use to acquire (and analyze) data and could realize that they are able to build their own acquisition software.

Although designing experiments was ranked as at least somewhat important by the majority of surveyed instructors, it was only discussed by a few instructors when they were asked broadly about important experimental skills in the interviews. Instructor 6  wanted students to learn about  ``experimental design in ways [the students] don't realize they're learning it'' 
because the students are learning how to set up an experiment with all the small decisions such as where to place mirrors and irises. The amount of experimental design may depend on how the courses implement these experiments. For example, one instructor said there was less experimental design with the single-photon experiments than other experiments in their course because the students were provided a manual to follow.

There were other general lab skills brought up in the interviews as well. One instructor discussed how the students learned to keep themselves safe from the lasers and to keep the equipment safe from the students (e.g., not turning on the room lights when the avalanche photodiodes were on). Other instructors mentioned the goal of learning how to use standard lab equipment such as oscilloscopes, spectrometers, and multimeters. Some instructors focused on the troubleshooting skills that come from working with an experiment. Instructors wanted students to be able to understand the entire process of the experiment, to be comfortable playing around with the equipment, and to learn how to solve problems they did not know how to solve initially.

\subsection{Professional skills}

One theme that appeared in the interviews was the idea that the single-photon experiments may help students learn how to be persistent and keep trying even after failed initial attempts. Although it was rarely explicitly taught, some instructors wanted students to learn how to deal with frustration because it would help them later in life, whether that was graduate school or anywhere else. This sense of frustration may be particularly relevant for the single-photon experiments because they involve very precise optical alignment where it can take weeks before a signal is first detected. There are many aspects of a project that could lead to frustration, and one instructor discussed ways they try to balance student frustration across different projects in their course by providing more guidance (i.e., a clear set of instructions or manual) for the more technically demanding experiments (such as the single-photon experiments) and allowing students to make more decisions for the less technically demanding ones.

Decision making and independence was another skill some instructors hoped students would gain while working with the single-photon experiments. For that aim, some of the instructors provided students papers or other resources, and expected the students to read them and make their own decisions from there. The decisions could include choices about which experimental question to investigate or how to make the measurement. It was important to some instructors to encourage their students to take ownership over the experiment.

Another set of skills that was important for many of the instructors was the communication and presentation of results. Different instructors focused on different aspects that are commonly performed by practicing physicists. Some instructors hoped to use the single-photon experiments to teach students how to keep good lab notebooks, write lab reports or research articles, or read journal articles. Others wanted to help their students more broadly be able to discuss technical topics. When asked to elaborate on the stated goal of communication, Instructor 2 explained how they used this experiment (and the other experiments offered in their courses) as an opportunity for students to give short presentations about their work multiple times throughout the course:
\begin{quote} 
    \textit{``I want them to be able to speak on a technical topic without a lot of fluff... When they start off, usually they spend their five minutes not saying a whole lot... And so my hope is, by the end, they will have a better sense as to how to put together a talk that speaks to a specific question, says what needs to be said, and doesn't have a lot of terms related to how really incredibly awesomely exciting this experiment is and... all of those things that students do to fill the time. My goal is that once they go off to a job interview or once they are in grad school doing their quals or whatever, they'll have gleaned a little bit of comfort speaking on a technical topic.''}
\end{quote}
Working with complex experiments can provide students the opportunity to practice a variety of authentic communication methods.

\subsection{Epistemological views}

The last category of goals that were brought up in the interviews relates to the importance of experimentation and the connection between theory and experiment. Because the single-photon experiments are about foundational concepts in quantum mechanics that can lead to a fundamental shift in the way people perceive the world, they could be uniquely situated to help students grapple with epistemological ideas. Some instructors mentioned that they wanted students to understand how theory and experiment work together in physics. One instructor wanted students to learn that the canonical spin-1/2 system discussed in all quantum classes is something that you can actually do experiments with. It is more than just a mathematical construct; it is a tool to explain what happens in the world. The idea that these experiments help demonstrate that the theory of quantum mechanics accurately describes the physical world is connected to the goal of seeing quantum mechanics (Sec.~\ref{sec:seeingQM}). 

Other instructors focused on the idea that questions about the world are answered through experimentation. Instructor 3, when asked to explain their additional learning goal of students appreciating how successful quantum mechanics is at explaining physical phenomena, said: 
\begin{quote} 
    \textit{``I think they do see the difference between classical and quantum, and that to ultimately know, you've got to go do something in the lab and test your model and figure out a way that your experiment can distinguish between this and that.''}
\end{quote}
Instructor 5, when also asked about additional learning goals, compared the single-photon experiments with widely-used experiments in introductory classes where students often try to force their data to match a known value. They talked about how students being able to ``conceptualize their experiment as their own experiment'' can help them learn to draw their own conclusions from data. 

\bibliography{singlePhotonExperiments}

\end{document}